\documentclass[sigconf]{acmart}
\AtBeginDocument{%
  }

\renewcommand\footnotetextcopyrightpermission[1]{}

\setcopyright{acmlicensed}
\copyrightyear{2018}
\acmYear{2018}
\acmDOI{XXXXXXX.XXXXXXX}
\acmConference[Conference acronym 'XX]{Make sure to enter the correct
  conference title from your rights confirmation email}{June 03--05,
  2018}{Woodstock, NY}
\acmISBN{978-1-4503-XXXX-X/2018/06}

\usepackage{subcaption}
\usepackage{multirow}
\usepackage{enumitem}
\usepackage{graphicx}
\usepackage{balance}
\usepackage{amsmath}
\usepackage{booktabs}
\usepackage{amssymb}
\usepackage{pifont}

\usepackage{amsmath}
\usepackage{algorithm}
\usepackage{algpseudocode}
\usepackage{amsfonts}

\usepackage{listings}
\usepackage{url}
\usepackage{hyperref}

\usepackage{colortbl}  
\usepackage{xcolor}
\usepackage{array}   
\usepackage{marvosym} 

\definecolor{Highlight}{rgb}{0.94, 0.97, 1.0}

\newcommand{\best}[1]{\textbf{#1}}
\newcommand{\second}[1]{\underline{#1}}

\definecolor{linecolor}{gray}{.91} % soft gray
\definecolor{linecolor2}{gray}{.95} % soft gray
\definecolor{linecolor1}{gray}{.97} % soft gray

\usepackage{color, xspace}
\usepackage[normalem]{ulem}

\lstdefinestyle{custom_style}{
    basicstyle=\ttfamily\small,     
    breaklines=true,                 
    frame=single,                    
    xleftmargin=1pt,               
    xrightmargin=1pt,               
    keywordstyle=\color{blue},       
    commentstyle=\color{gray},      
    stringstyle=\color{red},         
}
\usepackage[capitalize]{cleveref}

\begin{document}

%%
%% The "title" command has an optional parameter,
%% allowing the author to define a "short title" to be used in page headers.

\title{PRISM: Purified Representation and Integrated Semantic Modeling for Generative Sequential Recommendation}

\author{Dengzhao Fang}
\affiliation{%
  \institution{Jilin University}
    \city{Changchun}
  \country{China}
}
\email{fangdz25@mails.jlu.edu.cn}

\author{Jingtong Gao}
\affiliation{%
  \institution{City University of Hong Kong}
    \city{Hong Kong}
  \country{China}
}
\email{jt.g@my.cityu.edu.hk}

\author{Yu Li}
\authornote{Corresponding author.}
\affiliation{%
  \institution{Jilin University}
    \city{Changchun}
  \country{China}
}
\email{liyu90@jlu.edu.cn}

\author{Xiangyu Zhao}
\affiliation{%
  \institution{City University of Hong Kong}
    \city{Hong Kong}
  \country{China}
}
\email{xianzhao@cityu.edu.hk}

\author{Yi Chang}
\affiliation{%
  \institution{Jilin University}
    \city{Changchun}
  \country{China}
}
\email{yichang@jlu.edu.cn}

%%
%% The "author" command and its associated commands are used to define
%% the authors and their affiliations.
%% Of note is the shared affiliation of the first two authors, and the
%% "authornote" and "authornotemark" commands
%% used to denote shared contribution to the research.

%%
%% By default, the full list of authors will be used in the page
%% headers. Often, this list is too long, and will overlap
%% other information printed in the page headers. This command allows
%% the author to define a more concise list
%% of authors' names for this purpose.
\renewcommand{\shortauthors}{Trovato et al.}

\begin{abstract}

Generative Sequential Recommendation (GSR) has emerged as a promising paradigm, reframing recommendation as an autoregressive sequence generation task over discrete Semantic IDs (SIDs), typically derived via codebook-based quantization.
Despite its great potential in unifying retrieval and ranking, existing GSR frameworks still face two critical limitations: (1)~\textit{impure and unstable semantic tokenization}, where quantization methods struggle with interaction noise and codebook collapse, resulting in SIDs with ambiguous discrimination; and (2)~\textit{lossy and weakly structured generation}, where reliance solely on coarse-grained discrete tokens inevitably introduces information loss and neglects items' hierarchical logic.
To address these issues, we propose a novel generative recommendation framework, PRISM, with Purified Representation and Integrated Semantic Modeling. Specifically, to ensure high-quality tokenization, we design a Purified Semantic Quantizer that constructs a robust codebook via adaptive collaborative denoising and hierarchical semantic anchoring mechanisms. To compensate for information loss during quantization, we further propose an Integrated Semantic Recommender, which incorporates a dynamic semantic integration mechanism to fuse fine-grained semantics and enforces logical validity through a semantic structure alignment objective. PRISM consistently outperforms state-of-the-art baselines across four real-world datasets, demonstrating substantial performance gains, particularly in high-sparsity scenarios. 
\end{abstract}

%%
%% The code below is generated by the tool at http://dl.acm.org/ccs.cfm.
%% Please copy and paste the code instead of the example below.
%%

%%
%% Keywords. The author(s) should pick words that accurately describe
%% the work being presented. Separate the keywords with commas.
%%
%% This command processes the author and affiliation and title
%% information and builds the first part of the formatted document.
\maketitle

\section{Introduction}
\label{sec:introduction}
Sequential recommender systems (SRSs) aim to predict a user's future interests based on interaction history~\cite{zhang2019deep, chen2021modeling, LinRec, SMLP4Rec}. Discriminative models like SASRec~\cite{SASRec} and BERT4Rec~\cite{BERT4Rec} have advanced SRSs by modeling dynamic dependencies. These models typically adhere to a ``retrieve-and-rank'' paradigm, relying on external similarity retrieval~\cite{Faiss} over collaborative embeddings.
Despite their effectiveness, this paradigm suffers from inherent limitations: the disjointed retrieval-and-rank architecture leads to objective misalignment, hindering global optimization, while the reliance on random atomic IDs neglects intrinsic item semantics, restricting the generalizability and expressiveness of item representations.

As a promising alternative, Generative Sequential Recommendation (GSR), inspired by Large Language Models (LLMs)~\cite{zhang2023instruction, wu2024survey, zhang2025killing}, has emerged as a new paradigm that reframes recommendation as an autoregressive sequence generation~\cite{deng2025onerec, li2025semantic, ETEGRec, BBQRec, DiscRec}. Unlike discriminative models that treat items as independent labels, GSR explicitly leverages the semantic correlations within item content.
Its core concept is that, during the \textbf{semantic tokenization} stage, each item is represented as a sequence of discrete ``Semantic IDs'' (SIDs)~\cite{TIGER, GRID, hua2023index, ActionPiece}, which are derived from semantic features, e.g., textual descriptions, in contrast to the atomic IDs used in discriminative models.
Subsequently, in the \textbf{generative recommendation} stage, a Transformer~\cite{vaswani2017attention} model is used to generate the SIDs of the next item autoregressively. This paradigm addresses discriminative limitations by eliminating structural fragmentation and enabling knowledge transfer across items sharing similar semantic codes, thereby empowering the model to capture deeper user intents and improving items' representation robustness.

From the perspective of the recommendation backbone, existing GSR paradigms can be broadly categorized into two streams. 
The first stream leverages LLMs~\cite{LC-Rec, ColaRec, Llara, HLLM, Ye2025Align3GRUM}, which typically construct a codebook and subsequently employ instruction-tuning to align SIDs with textual knowledge. 
The second stream focuses on lightweight models~\cite{TIGER, EAGER, LETTER, ActionPiece} that construct codebooks and then learn sequence generation from scratch. 
Despite the superior semantic reasoning capability of LLM-based methods, they suffer from prohibitive training costs and high inference latency~\cite{zhou2025efficiency, xi2025efficiency}, making them difficult for real-time deployment. 
Consequently, \textit{in this work, we specifically focus on lightweight generative frameworks}, which offer a pragmatic trade-off between efficiency and effectiveness. 
While promising, current lightweight frameworks face two critical limitations that hinder their full potential.

\begin{figure}[t]
    \centering
    % \vspace{-2mm}
    \includegraphics[width=0.8\linewidth]{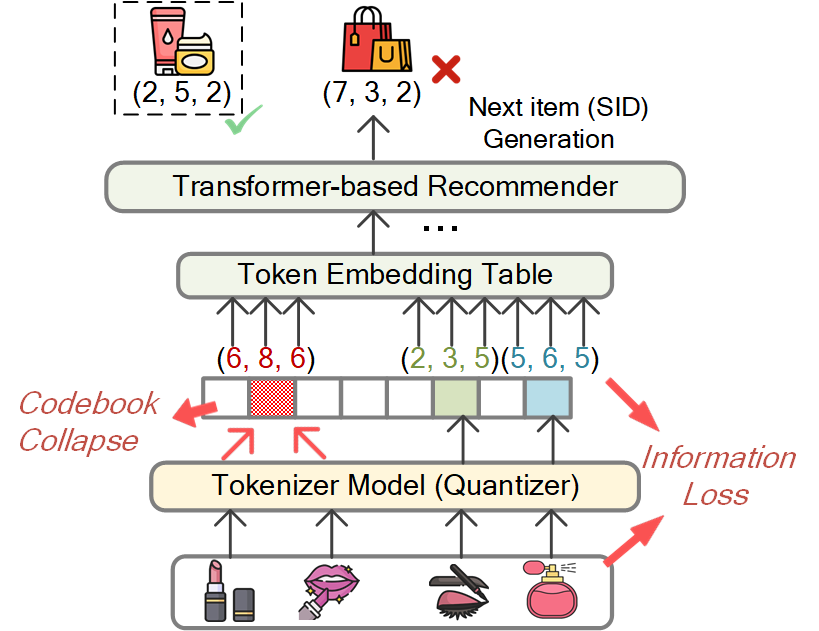}    
    \vspace{-3pt}
    \Description{A diagram illustrating two critical limitations in existing Generative Sequential Recommendation frameworks. The top section shows a Transformer-based recommender generating semantic IDs. The middle section highlights `Codebook Collapse', depicting a tokenizer mapping diverse items to the same few red-colored codes in the embedding table, causing indistinguishability. The bottom section highlights `Information Loss', showing that discrete Semantic IDs fail to capture the fine-grained continuous features from the original item images.}
    \caption{
    \textbf{Illustration of critical limitations in existing GSR frameworks.} 
    (a) \textbf{Codebook Collapse:} The unstable quantizer tokenizes diverse items into a narrow range of codes, making items indistinguishable. 
    (b) \textbf{Information Loss:} Discrete SIDs fail to capture fine-grained continuous semantics, providing insufficient item features for recommendation. }
    \label{fig:intro}
    \vspace{-11pt}
\end{figure}

\textit{\textbf{Impure and unstable semantic tokenization.}} 
Constructing a high-quality discrete SID vocabulary, i.e., codebook, is the cornerstone of GSR, yet existing methods struggle to balance semantic expressiveness with structural stability. 
%%%%
Current methods often ignore collaborative signals~\cite{TIGER, LC-Rec, RPG, TokenRec} or rely on rigid clustering~\cite{deng2025onerec, PRORec}, which inherently fails to provide distinct identifiers for items. 
Although recent learnable quantization methods~\cite{ColaRec, LETTER, EAGER, HLLM} attempt to integrate content with collaborative signals, their direct fusion of collaborative features introduces interaction noise~\cite{SimGCL, SGL, LLM4DSR}, thereby obscuring the fine-grained distinctions between items. 
Moreover, the quantization optimization itself is unstable, frequently leading to \textit{codebook collapse}~\cite{van2017neural,kuai-etal-2024-breaking}, where a majority of the vocabulary remains unused, as illustrated in Figure~\ref{fig:intro}(a). 
Consequently, the learned SIDs become semantically ambiguous and insufficiently discriminative, failing to index complex user intents accurately.

\textit{\textbf{Lossy and weakly structured generation.}} 
The second limitation lies in the utilization of discrete SIDs during generation. Unlike pre-trained LLMs that can rely on vast internal knowledge to bridge semantic gaps~\cite{LC-Rec, ColaRec, Llara, HLLM}, lightweight models structurally isolate the generation process from the continuous feature space. 
As depicted in Figure~\ref{fig:intro}(b), relying solely on discrete SIDs causes severe \textit{information loss}, as fine-grained content details and collaborative signals are discarded after quantization, rendering the model incapable of distinguishing items with subtle semantic differences.
Furthermore, generating flat SIDs overlooks the intrinsic categorical logic, resulting in structural misalignment where generated tokens may be numerically valid but semantically deviant.

In essence, lightweight frameworks struggle to maintain a discriminative index during tokenization, while suffering severe information loss during generation.
Given these limitations, \textit{how can we construct a robust semantic index while simultaneously compensating for information loss during lightweight generation?} To answer this question, we propose \textbf{PRISM} (\underline{P}urified \underline{R}epresentation and \underline{I}ntegrated \underline{S}emantic \underline{M}odeling), a unified framework that synergizes purified structural tokenization and dynamic semantic integration.
Specifically, to construct a robust codebook, we design a \textbf{Purified Semantic Quantizer}. It introduces an adaptive purification mechanism that distills collaborative signals from interaction noise, while simultaneously enforcing hierarchical constraints to anchor SIDs to their intrinsic categories. This ensures the learned codebook is both discriminative and structurally robust against collapse.
Furthermore, to address information loss without sacrificing the structural efficiency of discrete SIDs, we propose an \textbf{Integrated Semantic Recommender}.
Specifically, this module compensates for the information loss by dynamically integrating continuous semantic features into the discrete autoregressive generation process.
This design allows PRISM to enjoy the best of both worlds, retaining the efficient indexability of discrete tokens while recovering the fine-grained fidelity of continuous representations.
Finally, to ensure the generated sequences are logically valid, we align the generation process with the item's hierarchical structure. 
The main contributions are summarized as follows:
\begin{itemize}[leftmargin=*, nosep]

    \item We propose PRISM, a novel framework that synergizes signal purification in tokenization and dynamic semantic integration in generation, effectively addressing the twin limitations of codebook collapse and quantization information loss.
    
    \item We propose a \textit{Purified Semantic Quantizer} that constructs a robust codebook by distilling noisy signals and imposing hierarchical constraints, ensuring both signal purity and structural stability.

    \item We introduce an \textit{Integrated Semantic Recommender} that employs a dynamic mechanism to compensate for quantization loss, adaptively balancing different features to achieve precise generation.

     \item Extensive experiments on four datasets demonstrate that PRISM significantly outperforms state-of-the-art (SOTA) baselines, particularly achieving remarkable gains in sparse data scenarios.

\end{itemize}

\section{Related Work}
\label{sec:related}

\begin{figure*}[t!]
    \centering 
    \includegraphics[width=0.99\linewidth]{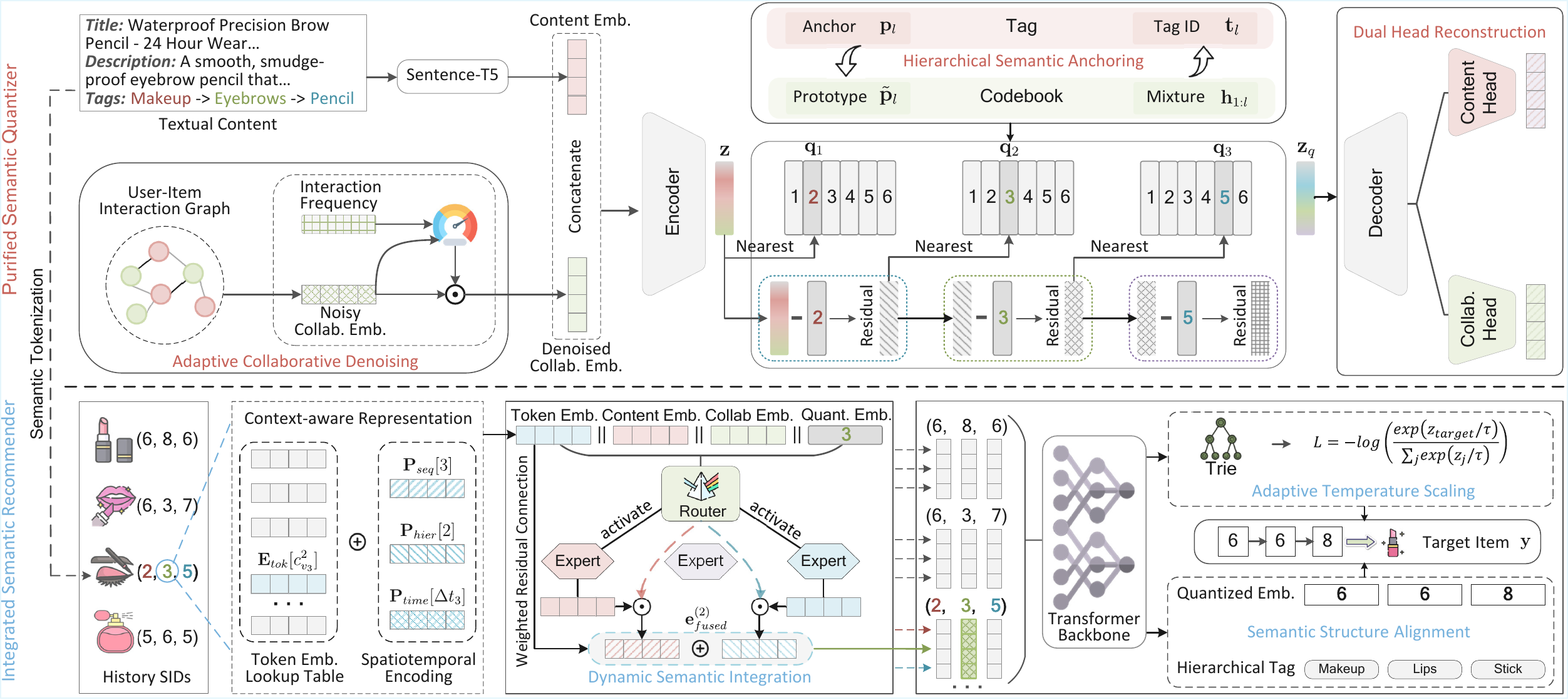}
    % \vspace{-8pt}
    \Description{The overall architecture of the PRISM framework, divided into two main stages. The top panel shows the Purified Semantic Quantizer, which processes textual content through a Sentence-T5 encoder and an interaction graph. It features Adaptive Collaborative Denoising to filter noise and Hierarchical Semantic Anchoring to align embeddings with category tags before quantization. The bottom panel shows the Integrated Semantic Recommender. It takes history SIDs, processes them through a context-aware representation layer, and uses a Dynamic Semantic Integration module with a Mixture-of-Experts router to fuse continuous and discrete features. The final output is generated via a Transformer backbone using Adaptive Temperature Scaling and is supervised by a Semantic Structure Alignment objective.}
    \caption{{\bf The PRISM framework.} 
    PRISM first learns a robust vocabulary via the Purified Semantic Quantizer, 
    and then the Integrated Semantic Recommender utilizes the vocabulary to tokenize items into semantic IDs for generative recommendation.
    % \ly{check the statement.}
    }
    \label{fig:framework}
    % \vspace{-10pt}
  \end{figure*}

% \subsection{Traditional Sequential Recommendation}
Modeling user preference evolution is central to recommendations. Early methods, such as GRU4Rec~\cite{GRU4Rec}, used RNNs to capture sequential patterns. The field later adopted Transformer~\cite{vaswani2017attention} to model long-range dependencies. Prominent models like SASRec~\cite{SASRec} use unidirectional self-attention, while BERT4Rec~\cite{BERT4Rec} applies bidirectional masking. Although some works like S$^3$-Rec~\cite{S3-rec} improve representations through pre-training, they remain discriminative. They separate representation learning from retrieval, requiring external indices~\cite{Faiss} to score candidates. Furthermore, the reliance on rigid atomic IDs neglects intrinsic item semantics, limiting expressiveness. These limitations prompt the shift to generative frameworks.

% \subsection{Generative Sequential Recommendation}
A recent paradigm shift reframes recommendation as an autoregressive sequence generation task named generative sequential recommendation (GSR), unifying retrieval and ranking into an end-to-end step~\cite{li2023e4srec,ji2024genrec,wu2024survey,li2024large,wang2023generative,zhang2025recommendation,lopez2025survey, wang2024recommendation}. While this paradigm promises greater flexibility, its success hinges on two interdependent challenges~\cite{liu2024vector,zhai2024actions,HLLM,jia2025principles,hou2025generative,li2025survey,li2025discrete}: the discriminative quality of the discrete item representations (SIDs)~\cite{TIGER,GRID,hua2023index} and the effectiveness of their end-to-end optimization during recommendation.
Below, we review existing GSR from two corresponding perspectives: semantic tokenization and generative recommendation.

\textbf{Semantic Tokenization.}
The performance of GSR fundamentally depends on the quality of SIDs, which are typically tokenized via either non-parametric clustering or learnable quantization. Learnable methods, pioneered by TIGER~\cite{TIGER}, employ residual quantization networks~\cite{hou2023learning, RPG}, such as RQ-VAE~\cite{van2017neural, liang2018variational, RQ-VAE}, to map continuous item features into SIDs. This paradigm has evolved to integrate collaborative signals, as in LETTER~\cite{LETTER} and EAGER~\cite{EAGER}, which fuse content and collaborative features during quantization. Conversely, non-parametric methods like Residual K-means~\cite{deng2025onerec} utilize fixed clustering algorithms on static representations. More recently, ActionPiece~\cite{ActionPiece} introduces a context-aware tokenization strategy that merges frequent feature patterns based on co-occurrence, offering a dynamic alternative to static ID assignment. However, these methods face distinct limitations. Non-parametric methods lack the flexibility to fuse content and collaborative features~\cite{EAGER, PRORec}. Meanwhile, learnable methods suffer from optimization instabilities like codebook collapse and index collisions~\cite{LC-Rec,kuai-etal-2024-breaking,deng2025onerec,hidvae}. Furthermore, incorporating collaborative signals introduces interaction noise that often corrupts the codebook structure~\cite{SimGCL, SGL, UDT}, thereby obscuring item semantics and degrading recommendation performance. Different from them, our PRISM constructs a robust and purified vocabulary by adaptively filtering interaction noise and explicitly enforcing hierarchical structural stability.

\textbf{Generative Recommendation.}
Generative recommendation reformulates the task as an end-to-end autoregressive generation over discrete SIDs. Existing frameworks fall into two categories based on their backbone architectures. The first category leverages pre-trained LLMs to mitigate the inherent information loss of quantization~\cite{jin2024language,geng2022recommendation,huangimproving,pang2025generative}. Methods like LC-Rec~\cite{LC-Rec} and ColaRec~\cite{ColaRec} employ instruction tuning to inject textual semantics, while HLLM~\cite{HLLM}, LLARA~\cite{Llara}, and COBRA~\cite{yang2025sparse} align dense features with SIDs within large-scale backbones. Despite their superior semantic understanding, the massive parameter scale of LLMs imposes prohibitive latency and computational cost, rendering them impractical for real-time retrieval~\cite{zhou2025efficiency, xi2025efficiency, Lin2025OrderagnosticIF}. Consequently, the second category focuses on lightweight generative frameworks to prioritize efficiency~\cite{ETEGRec, BBQRec, DiscRec, RPG, PRORec, zhu2025adaptive}. Representative methods, such as TIGER~\cite{TIGER}, LETTER~\cite{LETTER}, EAGER~\cite{EAGER}, and ActionPiece~\cite{ActionPiece}, employ compact Transformers to predict SIDs. However, these methods primarily focus on infusing heterogeneous signals into SIDs during the tokenization stage, neglecting to dynamically leverage continuous features during the generative stage to compensate for quantization loss. Moreover, most of them treat hierarchical SIDs as flat sequences, disregarding the intrinsic structural logic of the identifier tree. In contrast, our PRISM compensates for the information loss caused by quantization by dynamically integrating fine-grained continuous signals and ensuring logical validity through semantic structure alignment.

\section{Methodology}
\label{sec:method}

\subsection{Problem Formulation}
Let $\mathcal{U}$ and $\mathcal{I}$ denote the sets of users and items. For each user $u \in \mathcal{U}$, the interaction history is denoted as a sequence $\boldsymbol{S}_u = [v_1, v_2, \dots, v_t]$. The objective is to predict the next item $v_{t+1}$.
Instead of atomic IDs, we represent each item as a sequence of discrete SIDs. Specifically, we employ a quantizer $Q(\cdot)$ to encode each item $v \in \mathcal{I}$ into a token sequence $\boldsymbol{c}_{v} = (c_{v}^{1}, \dots, c_{v}^{L})$ of length $L$, where each token $c_{v}^{l}$ belongs to a vocabulary $\mathcal{V}$. This transforms the recommendation task into an autoregressive sequence generation problem.

Formally, let $\mathbf{y} = (y_1, \dots, y_L) = \boldsymbol{c}_{v_{t+1}}$ denote the discrete token sequence of the target item $v_{t+1}$. Given the history $\boldsymbol{S}_u$, which is tokenized element-wise as $Q(\boldsymbol{S}_u)$, we formulate the next-item prediction as an autoregressive generation task. The probability of generating the target item $\mathbf{y}$ is decomposed via the chain rule:
\begin{equation}
    p(\mathbf{y} \mid \boldsymbol{S}_u) = \prod_{l=1}^{L} p(y_l \mid Q(\boldsymbol{S}_u), \mathbf{y}_{<l}),
    \label{eq:quantizer}
\end{equation}
where $y_l$ denotes the $l$-th token of the target sequence, and $\mathbf{y}_{<l} = (y_1, \dots, y_{l-1})$ denotes the preceding generated tokens.

\subsection{Overall Pipeline} 

As illustrated in Figure~\ref{fig:framework}, PRISM bridges semantic tokenization and generative recommendation through a unified two-stage framework. In the first stage, the Purified Semantic Quantizer constructs a robust SIDs vocabulary with signal purity and codebook stability. In the second stage, the Integrated Semantic Recommender performs autoregressive generation while dynamically integrating continuous features to compensate for quantization loss, ensuring both accurate and structurally consistent recommendations.

\subsection{Purified Semantic Quantizer}

\subsubsection{Adaptive Collaborative Denoising for Signal Purification}
\label{sec:acd}

To construct robust SIDs, it is essential to integrate collaborative signals with content features. However, directly fusing these heterogeneous signals poses a risk: interaction data usually contains noise that can contaminate the encoder. To address this, we propose an Adaptive Collaborative Denoising (ACD) mechanism to filter unreliable patterns before fusion. Intuitively, we expect the quantizer to prioritize collaborative signals for popular items while moving to content features for long-tail items to mitigate noise interference.

Formally, for an item $v$, let $\mathbf{e}_{cont} \in \mathbb{R}^{d_{cont}}$ and $\mathbf{e}_{collab} \in \mathbb{R}^{d_{collab}}$ be its pre-extracted content and collaborative embeddings.\footnote{In our implementation, $\mathbf{e}_{cont}$ ($d_{cont}=768$) is from \texttt{sentence-t5-base}~\cite{ni2022sentence}, and $\mathbf{e}_{collab}$ ($d_{collab}=64$) is from \texttt{LightGCN}\cite{LightGCN}.} Our goal is to generate an element-wise trust gate $\mathbf{g} \in (0, 1)^{d_{collab}}$ to selectively retain collaborative features.
Specifically, we learn the trust gate vector $\mathbf{g}$ through a learnable gating network $\phi_{gate}(\cdot)$, which is parameterized as a Multi-Layer Perceptron with a non-linear activation in the hidden layer. 
\begin{equation}
\mathbf{g} = \sigma(\phi_{gate}(\mathbf{e}_{collab})),
\end{equation}
where $\sigma(\cdot)$ denotes the Sigmoid activation function. Subsequently, we can obtain the purified collaborative embedding:
\begin{equation}
    \tilde{\mathbf{e}}_{collab} = \mathbf{g} \odot \mathbf{e}_{collab},
\end{equation}
where $\odot$ denotes element-wise multiplication.

However, relying solely on implicit backpropagation to optimize the trust gate $\mathbf{g}$ often leads to convergence issues. To ensure that the gating vector $\mathbf{g}$ can effectively enhance the reliability of noisy signal identification and to prevent all gating values from collapsing into a single scalar, we use item interaction frequency as an empirical proxy for reliability, serving as auxiliary supervision. This is based on the observation that popular items typically exhibit more stable collaborative patterns compared to sparse items~\cite{SimGCL,SGL}. Specifically, we align the average gating values with item popularity through the following loss function. 
\begin{equation}
\mathcal{L}_{acd} = (\bar{g}_v - p_v)^2 + \mathrm{ReLU}(\delta - \mathrm{Var}(\mathbf{g}_v)),
\end{equation}
where $p_v$ is the popularity of item $v$ normalized to $[0, 1]$ via Min-Max scaling, $\bar{g}_v$ denotes the mean value of the gating vector $\mathbf{g}_v$ across its feature dimensions. The second term serves as a diversity regularizer, where $\mathrm{Var}(\mathbf{g}_v)$ computes the variance of the gate elements, and $\delta$ is a margin hyperparameter.

\subsubsection{Hierarchical Semantic Anchoring for Latent Stability}
\label{sec:hsa}

To generate discrete SIDs from the heterogeneous signals, we adopt RQ-VAE~\cite{zeghidour2022soundstream, RQ-VAE} to instantiate the quantization function $Q(\cdot)$ defined in Eq.~\ref{eq:quantizer}.
We fuse the content and purified collaborative embeddings via an encoder to obtain the continuous latent representation $\mathbf{z}$:
\begin{equation}
    \mathbf{z} = \mathrm{Enc}(\mathbf{e}_{cont} \parallel \tilde{\mathbf{e}}_{collab}),
\end{equation}
where $\parallel$ denotes concatenation and $\mathrm{Enc}(\cdot)$ is the RQ-VAE encoder.

However, performing residual quantization on $\mathbf{z}$ with randomly initialized codebooks lacks semantic guidance. This unconstrained recursive approximation often leads to severe codebook collapse due to optimization instability. Therefore, we propose a Hierarchical Semantic Anchoring (HSA) module, which leverages hierarchical category tags, such as {``Makeup-Eyebrows-Pencil''}, as semantic priors to organize the codebook, thereby mirroring the coarse-to-fine nature of residual quantization.\footnote{Category tags are intrinsic to standard benchmarks~\cite{liu2025cat}. Even if absent, reliable hierarchies can be robustly synthesized via LLMs~\cite{tang2025llm4tag}.}

For the codebook at any layer $l$, where $1\leq l\leq L$, HSA imposes a dual alignment constraint. 
In the first alignment, we encode the tag corresponding to depth $l$ using the same backbone as $\mathbf{e}_{cont}$ and linearly project it into the codebook space to serve as the semantic anchor $\mathbf{p}_l$.
To avoid rigid constraints, we construct a soft prototype embedding $\tilde{\mathbf{p}}_l$ by aggregating candidate embeddings from the layer-specific codebook $\mathcal{C}_l$, weighted by their proximity to the anchor:
\begin{equation}\label{eq:soft_quantized}
    \tilde{\mathbf{p}}_l = \sum_{k=1}^{|\mathcal{C}_l|} \alpha_k \mathbf{e}_k^{(l)}, \;\; \alpha_k = \frac{\exp(-\|\mathbf{p}_l - \mathbf{e}_k^{(l)}\|^2 / \tau_{hsa})}{\sum_{j=1}^{|\mathcal{C}_l|} \exp(-\|\mathbf{p}_l - \mathbf{e}_j^{(l)}\|^2 / \tau_{hsa})},
\end{equation}
where $\mathcal{C}_l$ is the $l$-th codebook, $\alpha_k$ is the attention weight for the $k$-th codebook embedding $\mathbf{e}_k^{(l)}$, and $\tau_{hsa}$ regulates distribution sharpness.

The second alignment ensures semantic preservation through classification-based supervision. Given the quantized embedding $\mathbf{q}_l \in \mathcal{C}_l$, i.e., the closest codebook embedding selected by the quantizer at depth $l$, we use the mixed representation $\mathbf{h}_{1:l} = \mathbf{q}_1 \parallel \dots \parallel \mathbf{q}_l$ to predict the ground-truth tag $\mathbf{t}_l$ at depth $l$, where this mixed representation is obtained by concatenating quantized embedding from depth 1 to $l$. The dual alignment constraint is optimized as: 
\begin{equation}
    \mathcal{L}_{hsa} = \sum_{l=1}^{L} \left( \|\mathbf{p}_l - \tilde{\mathbf{p}}_l\|^2 + \text{CE}(\phi_{cls}^{(l)}(\mathbf{h}_{1:l}), \mathbf{t}_l) \right),
\end{equation}
where $\text{CE}(\cdot)$ denotes the Cross-Entropy loss, and $\phi_{cls}^{(l)}(\cdot)$ is the linear classifier used for tag prediction.

\subsubsection{Dual-Head Reconstruction and Optimization}
\label{sec:joint_opt}

To drive the learning of the encoder and the hierarchical codebooks structured by HSA, we employ a reconstruction objective. Following the residual quantization paradigm, we compute the final quantized representation $\mathbf{z}_q$ by summing the quantized embeddings across all $L$ layers: $\mathbf{z}_q = \sum_{l=1}^L \mathbf{q}_l$. This aggregated representation is then used to reconstruct the original input features.

However, a critical issue stems from gradient imbalance, where high-dimensional content embeddings dominate the optimization over collaborative signals~\cite{Peng2022BalancedML, Yuan2023WhereTG}.  Consequently, the reconstruction decoder tends to neglect collaborative patterns. To resolve this, we propose a Dual-Head Reconstruction (DHR) objective, which enforces balanced supervision via task-specific decoders: 
\begin{equation}\label{eq:hdr}
    \mathcal{L}_{dhr} = \|\mathbf{e}_{cont} - \mathrm{Dec}_{cont}(\mathbf{z}_q)\|^2 + \|\tilde{\mathbf{e}}_{collab} - \mathrm{Dec}_{collab}(\mathbf{z}_q)\|^2,
\end{equation}
where $\mathrm{Dec}_{cont}(\cdot)$ and $\mathrm{Dec}_{collab}(\cdot)$ denote decoders that reconstruct $\mathbf{z}_q$ back to the content and purified collaborative spaces.

\textbf{Optimization Strategy.}
Since the quantization process involves a non-differentiable $\text{argmin}$ operation, we employ the Straight-Through Estimator (STE)~\cite{STE} to enable gradient backpropagation. Specifically, gradients flow directly from the decoder to the encoder, thereby bypassing the quantization bottleneck. 
For updating the codebook, we adopt the standard Exponential Moving Average (EMA)~\cite{van2017neural} as a momentum-based mechanism to maintain training stability. 
It is worth noting that although EMA smooths the numerical update process, the semantic structure of the codebook is still jointly determined by HSA and the reconstruction objective. 
Formally, the Purified Semantic Quantizer optimizes as:
\begin{equation}
    \mathcal{L}_{psq} = \mathcal{L}_{dhr} + \beta \mathcal{L}_{commit} + \lambda_1 \mathcal{L}_{acd} + \lambda_2 \mathcal{L}_{hsa},
\end{equation}
where $\mathcal{L}_{commit} = \|\mathbf{z} - \text{sg}[\mathbf{z}_q]\|^2$ aligns encoder outputs with the quantized space to support STE, $\text{sg}[\cdot]$ denotes the stop-gradient operator~\cite{van2017neural}, and $\beta, \lambda_1, \lambda_2$ are hyperparameters balancing the terms.

\textbf{Global Collision Deduplication.}
After the training of the Purified Semantic Quantizer converges, to resolve inevitable ID collisions that arise when distinct items map to identical SIDs, we perform a global deduplication procedure. Unlike heuristic methods that append numeric suffixes and thus disrupt semantics~\cite{TIGER}, we formulate this as an optimal transport problem and apply the Sinkhorn-Knopp algorithm~\cite{Cuturi2013SinkhornDL} to optimally redistribute colliding items to their nearest available unique SIDs. This guarantees a collision-free mapping while preserving the global semantic structure learned in the semantic tokenization stage.

\subsection{Integrated Semantic Recommender}

\subsubsection{Dynamic Semantic Integration via Mixture-of-Experts}
\label{sec:moe}

Although the semantic tokenization stage constructs a robust SIDs vocabulary, quantization is a form of lossy compression. If the subsequent generative recommendation relies solely on the discrete SIDs, it inevitably loses fine-grained information embedded in the continuous space, resulting in information loss~\cite{RQ-VAE, TIGER}. To address this issue, we propose a Dynamic Semantic Integration (DSI) mechanism based on the Mixture-of-Experts (MoE) architecture~\cite{shazeer2017,zhang2025hierarchical}.

To construct a context-aware embedding capable of capturing the multi-dimensional dynamics of user interactions, we enhance SIDs with comprehensive spatiotemporal encodings. Formally, given the $l$-th token $c_{v_i}^l$ of the $i$-th item in the interaction sequence, we compute the enhanced embedding $\tilde{\mathbf{e}}_{id}^{(l)}$ via element-wise addition:
\begin{equation}
    \tilde{\mathbf{e}}_{id}^{(l)} = \mathbf{E}_{tok}[c_{v_i}^l] + \mathbf{P}_{seq}[i] + \mathbf{P}_{hier}[l] + \mathbf{P}_{time}[\Delta t_i],
\end{equation}
where $\mathbf{E}_{tok}[c_{v_i}^l]$ denotes the learnable token embedding, $\mathbf{P}_{seq}[i]$ encodes the sequential position, $\mathbf{P}_{hier}[l]$ identifies the granular depth $l$, and $\mathbf{P}_{time}[\Delta t_i]$ models the time interval $\Delta t_i$. 
For brevity, we omit the item index $i$ hereafter.

To restore fine-grained details, we fuse $\tilde{\mathbf{e}}_{id}^{(l)}$ with the item's content ($\mathbf{e}_{cont}$), collaborative embeddings ($\mathbf{e}_{collab}$), and the quantized codebook embedding $\mathbf{q_l}$ corresponding to the SID token at depth $l$ from the fixed codebook. 
However, a critical challenge arises from the misalignment between static item-level features and the hierarchical structure of the SIDs. 
Since a single item corresponds to a sequence of tokens of length $L$, the common practice of broadcasting static embeddings ($\mathbf{e}_{cont}, \mathbf{e}_{collab}$) to every token ignores the coarse-to-fine semantic structure, where a token at depth $l$ only requires information at the corresponding granularity level.
To address this, we design depth-specific projections that adaptively align static signals to the semantic level associated with depth $l$. In addition, to prevent high-dimensional content modality from overwhelming the sparse collaborative signals, we project $\mathbf{e}_{cont}$ into a lower-dimensional space while applying a shallow projection to $\mathbf{e}_{collab}$ to align it with the SID space. 
% For any item in the interaction history, its composite embedding at depth $l$, denoted by $\mathbf{x}_l$, is defined as follows:
The composite embedding $\mathbf{x}_l$ of each item at depth $l$ is given by:
\begin{equation}
    \mathbf{x}_l = \tilde{\mathbf{e}}_{id}^{(l)} \parallel \phi_{cont}^{(l)}(\mathbf{e}_{cont}) \parallel \phi_{col}^{(l)}(\mathbf{e}_{collab}) \parallel \mathbf{q_l},
\end{equation}
where each $\phi_{*}^{(l)}(\cdot)$ denotes a depth-specific projector consisting of a linear transformation followed by Layer Normalization~\cite{layernorm}.

Then, $\mathbf{x}_l$ is fed into a MoE layer, which acts as a semantic router. To capture dynamic features, we employ $N$ expert networks $\{E_i\}_{i=1}^N$ with a gating network $G(\cdot)$. To ensure load balancing and efficiency, we implement a noisy top-$K$ routing strategy, where the gating scores are computed via a linear projection with injected noise: 
\begin{equation}
    G(\mathbf{x}_l) = \mathbf{W}_g \mathbf{x}_l + \boldsymbol{\epsilon},
\end{equation}
where $\mathbf{W}_g$ is a learnable weight matrix mapping the input $\mathbf{x}_l$ to $N$ expert logits, and $\boldsymbol{\epsilon} \sim \mathcal{N}(0, 1)$ is standard Gaussian noise.

We explicitly identify the subset $\mathcal{K}$ of active expert indices corresponding to the top-$K$ gating scores. The output of MoE is then derived as the specialized aggregation over these selected experts:
\begin{equation}
    \mathbf{h}_{moe}^{(l)} = \sum_{i \in \mathcal{K}} \frac{\exp(G(\mathbf{x}_l)_i)}{\sum_{j \in \mathcal{K}} \exp(G(\mathbf{x}_l)_j)} E_i(\mathbf{x}_l),
\end{equation}
where $E_i(\mathbf{x}_l)$ denotes the embedding learned by the $i$-th expert, and the fraction is the normalized routing weight over $\mathcal{K}$.

Finally, to inject details while preserving the raw structure, we fuse expert knowledge into $\tilde{\mathbf{e}}_{id}^{(l)}$ via a weighted residual connection:
\begin{equation}
    \mathbf{e}_{fused}^{(l)} = \tilde{\mathbf{e}}_{id}^{(l)} + \eta \cdot \phi_{out}(\mathbf{h}_{moe}^{(l)}),
\end{equation}
where $\phi_{out}(\cdot)$ projects the MoE output back to the dimension of ID embedding, and the learnable scalar $\eta$ regulates the fusion intensity.

\subsubsection{Semantic Structure Alignment for Generative Consistency}
\label{sec:ssa}

While the DSI module enriches input embeddings, the final generation step still faces a structural mismatch challenge. Standard autoregressive methods treat SIDs as flat labels, ignoring their hierarchical dependency and information loss caused by quantization. This often leads to generated SIDs that are numerically valid but semantically drifted. We thus propose a Semantic Structure Alignment (SSA) module, which enhances consistency through auxiliary structural regularization and a density-aware generative objective.

The fused embeddings derived from DSI constitute the interaction sequence input for the Transformer-based recommender. To autoregressively generate the target item, the decoder produces a sequence of states. Let $\mathbf{o}_l$ denote the decoder's final hidden state when generating the $l$-th token of the target item. The $\mathbf{o}_l$ aggregates the user's historical context and the partial SID tokens of the target item to predict the next SID token. To ensure that $\mathbf{o}_l$ remains aligned with the semantic structure of the target item, we introduce a multi-view regularization strategy.

To compensate for the information loss of discrete SIDs, we require $\mathbf{o}_l$ to regress the quantized codebook embedding $\mathbf{q_l}^{(tgt)}$ corresponding to the target item's ground-truth token at depth $l$. Simultaneously, to prevent semantic drift, we require $\mathbf{o}_l$ to predict the corresponding hierarchical tag $\mathbf{t_l}^{(tgt)}$. By jointly optimizing these two constraints, we ensure the generated SIDs preserve both fine-grained details and category logic:
\begin{equation}
    \mathcal{L}_{ssa} = \sum_{l=1}^{L} \left( \|\phi_{reg}^{(l)}(\mathbf{o}_l) - \mathbf{q_l}^{(tgt)}\|^2 + \text{CE}(\phi_{cls}^{(l)}(\mathbf{o}_l), \mathbf{t_l}^{(tgt)}) \right),
\end{equation}
where $\phi_{reg}^{(l)}(\cdot)$ and $\phi_{cls}^{(l)}(\cdot)$ are depth-specific projectors mapping $\mathbf{o}_l$ to the codebook latent and tag logit spaces, respectively.

\subsubsection{Adaptive Temperature Scaling Generation}
Building upon the structurally aligned $\mathbf{o}_l$, the primary task is to autoregressively predict the target SIDs. To ensure the validity of the generated results, we employ Trie-based constrained decoding~\cite{liu2025understanding, chan2025efficient}, which restricts the search space to valid child nodes. However, the branching density of the Trie is not uniform across different positions. In standard autoregressive generation, a static Softmax temperature $\tau$ fails to adapt to this heterogeneity, struggling to suppress hard negatives.

To address this, we propose an Adaptive Temperature Scaling (ATS) mechanism. Instead of using a fixed $\tau$, we introduce a density-dependent function $\tau_{gen}(\cdot)$ that dynamically computes a scalar temperature based on the branching factor $N_l$, defined as the number of valid child nodes retrieved from the pre-constructed Trie given the prefix at depth $l$. Specifically, to maintain discrimination in dense branches, we formulate $\tau_{gen}(N_l)$ as an exponential decay function:
\begin{equation}
    \tau_{gen}(N_l) = \tau_{\min} + (\tau_{\max} - \tau_{\min}) \cdot \exp\left(-\alpha \frac{N_l}{N_{ref}}\right),
\end{equation}
where $\tau_{\min}$ and $\tau_{\max}$ define the temperature range, $N_{ref}$ is a normalization constant that adapts the decay scale to the branching characteristics, and $\alpha$ controls the sensitivity. 

The density-aware generative objective is optimized as:
\begin{equation}
    \mathcal{L}_{gen} = -\sum_{l=1}^{L} \log\left( \frac{\exp(\mathbf{w}_{y_l}^\top \mathbf{o}_l / \tau_{gen}(N_l))} {\sum_{j \in \mathcal{V}_l} \exp(\mathbf{w}_j^\top \mathbf{o}_l / \tau_{gen}(N_l))} \right),
\end{equation}
where $y_l$ denotes the ground-truth SID token at depth $l$, $\mathcal{V}_l$ is the full vocabulary of SIDs at depth $l$, and $\mathbf{w}_j$ is the output embedding for token $j$. Note that training spans the full vocabulary, whereas inference is Trie-constrained.

Formally, the optimization objective of the Integrated Semantic Recommender can be formulated as follows:
\begin{equation}
    \mathcal{L}_{isr} = \mathcal{L}_{gen} + \gamma \mathcal{L}_{ssa},
\end{equation}
where $\gamma$ is a hyperparameter balancing the structural constraints.

\section{Experiments}
\label{sec:exp}

\subsection{Experimental Setup}

\noindent\textbf{Datasets.}
We evaluate on four diverse Amazon datasets~\cite{he2016ups}\footnote{Available at: \url{https://jmcauley.ucsd.edu/data/amazon/}}: ``Beauty'', ``Sports and Outdoors'', ``Toys and Games'', and ``CDs and Vinyl'', hereafter referred to as Beauty, Sports, Toys, and CDs. Statistics of these datasets are shown in Table~\ref{tab:dataset_statistics}. Adhering to established protocols~\cite{SASRec,BERT4Rec}, we apply 5-core filtering and set the maximum sequence length to 20, retaining only the most recent interactions.

\begin{table}[htbp]
    \centering
     \vspace{-8pt}
    \caption{Statistics of the datasets.}
    \label{tab:dataset_statistics}
    \vspace{-8pt}
    \setlength{\tabcolsep}{0.95mm}
    % \resizebox{\linewidth}{!}{
        \begin{tabular}{lrrrrr}
            \toprule
            \textbf{Dataset} & \textbf{\#Users} & \textbf{\#Items} & \textbf{\#Interactions} & \textbf{Avg. Len.} & \textbf{Sparsity} \\
            \midrule
            Beauty & 22,363 & 12,101 & 198,502 & 8.88 & 99.93\% \\
            Sports & 35,598 & 18,357 & 296,337 & 8.32 & 99.95\% \\
            Toys   & 19,412 & 11,924 & 167,597 & 8.63 & 99.93\% \\
            CDs    & 75,258 & 64,443 & 1,097,592 & 14.58 & 99.98\% \\
            \bottomrule
        \end{tabular}
    \vspace{-5pt}
\end{table}

\begin{table*}[htbp]

\centering

\caption{Overall performance comparison on four real datasets. Metrics are evaluated at Recall (R) and NDCG (N) @10 and @20. The \best{bold} and \second{underlined} values denote the best and runner-up results, respectively.}

\vspace{-5pt}

\label{tab:main_results}

\setlength{\tabcolsep}{1.0mm} 

% \vspace{-4mm}

\renewcommand{\arraystretch}{1}

% \resizebox{\textwidth}{!}{

\begin{tabular}{cl ccccccc ccccc}

\toprule

\multirow{2}{*}{Dataset} & \multirow{2}{*}{Metric} & \multicolumn{7}{c}{\emph{Traditional Models}} & \multicolumn{5}{c}{\emph{Generative Models}} \\

\cmidrule(lr){3-9} \cmidrule(lr){10-14}

& & GRU4Rec & Caser & HGN & NextItNet & LightGCN & SASRec & BERT4Rec & TIGER & LETTER & EAGER & ActionPiece$^\dagger$ & \cellcolor{Highlight}\textbf{PRISM} \\

\midrule

% --- Beauty ---

\multirow{4}{*}{Beauty}

& R@10 & 0.0253 & 0.0237 & 0.0194 & 0.0447 & 0.0438 & 0.0489 & 0.0413 & 0.0588 & 0.0616 & 0.0600 & \second{0.0667} & \cellcolor{Highlight}\best{0.0713} \\

& R@20 & 0.0426 & 0.0397 & 0.0309 & 0.0714 & 0.0690 & 0.0769 & 0.0627 & 0.0857 & 0.0940 & 0.0884 & \second{0.1013} & \cellcolor{Highlight}\best{0.1030} \\

& N@10 & 0.0121 & 0.0116 & 0.0093 & 0.0220 & 0.0212 & 0.0211 & 0.0220 & 0.0309 & 0.0335 & 0.0335 & \second{0.0345} & \cellcolor{Highlight}\best{0.0387} \\

& N@20 & 0.0164 & 0.0156 & 0.0122 & 0.0287 & 0.0276 & 0.0282 & 0.0274 & 0.0377 & 0.0416 & 0.0407 & \second{0.0432} & \cellcolor{Highlight}\best{0.0467} \\

\midrule

% --- Sports ---

\multirow{4}{*}{Sports}

& R@10 & 0.0192 & 0.0182 & 0.0122 & 0.0265 & 0.0279 & 0.0295 & 0.0203 & \second{0.0401} & 0.0391 & 0.0332 & 0.0231 & \cellcolor{Highlight}\best{0.0409} \\

& R@20 & 0.0296 & 0.0278 & 0.0202 & 0.0433 & 0.0473 & 0.0471 & 0.0348 & \second{0.0617} & 0.0597 & 0.0500 & 0.0401 & \cellcolor{Highlight}\best{0.0636} \\

& N@10 & 0.0101 & 0.0090 & 0.0067 & 0.0135 & 0.0143 & 0.0126 & 0.0101 & \best{0.0210} & \second{0.0206} & 0.0170 & 0.0103 & \cellcolor{Highlight}\second{0.0206} \\

& N@20 & 0.0127 & 0.0114 & 0.0087 & 0.0177 & 0.0192 & 0.0170 & 0.0137 & \best{0.0264} & 0.0258 & 0.0212 & 0.0145 & \cellcolor{Highlight}\best{0.0264} \\

\midrule

% --- Toys ---

\multirow{4}{*}{Toys}

& R@10 & 0.0179 & 0.0175 & 0.0132 & 0.0324 & 0.0430 & 0.0567 & 0.0354 & 0.0574 & 0.0527 & 0.0518 & \second{0.0623} & \cellcolor{Highlight}\best{0.0686} \\

& R@20 & 0.0319 & 0.0274 & 0.0227 & 0.0509 & 0.0633 & 0.0831 & 0.0518 & 0.0868 & 0.0808 & 0.0789 & \second{0.1010} & \cellcolor{Highlight}\best{0.1011} \\

& N@10 & 0.0086 & 0.0088 & 0.0068 & 0.0162 & 0.0214 & 0.0247 & 0.0186 & 0.0304 & 0.0269 & 0.0286 & \second{0.0313} & \cellcolor{Highlight}\best{0.0348} \\

& N@20 & 0.0121 & 0.0113 & 0.0092 & 0.0208 & 0.0265 & 0.0313 & 0.0227 & 0.0378 & 0.0340 & 0.0355 & \second{0.0410} & \cellcolor{Highlight}\best{0.0430} \\

\midrule

% --- CDs ---

\multirow{4}{*}{CDs}

& R@10 & 0.0377 & 0.0300 & 0.0057 & 0.0545 & 0.0518 & 0.0479 & 0.0566 & \second{0.0580} & 0.0515 & 0.0510 & 0.0552 & \cellcolor{Highlight}\best{0.0777} \\

& R@20 & 0.0629 & 0.0500 & 0.0096 & 0.0856 & 0.0798 & 0.0790 & 0.0870 & 0.0863 & 0.0765 & 0.0785 & \second{0.0912} & \cellcolor{Highlight}\best{0.1163} \\

& N@10 & 0.0186 & 0.0148 & 0.0029 & 0.0273 & 0.0262 & 0.0208 & 0.0285 & \second{0.0308} & 0.0273 & 0.0264 & 0.0276 & \cellcolor{Highlight}\best{0.0412} \\

& N@20 & 0.0250 & 0.0198 & 0.0039 & 0.0351 & 0.0332 & 0.0286 & 0.0361 & \second{0.0380} & 0.0336 & 0.0333 & 0.0366 & \cellcolor{Highlight}\best{0.0509} \\

\bottomrule

\multicolumn{14}{l}{

  \scriptsize

  $^\dagger$ ActionPiece uses a larger backbone on CDs (see Implementation Details). Under the unified backbone setting consistent with other baselines, it yields R@10/20: 0.0348/0.0573 and N@10/20: 0.0166/0.0223.

}

\end{tabular}

% }

% \vspace{-4mm}

\end{table*}

\noindent\textbf{Baselines.} We compare PRISM with two groups of SOTA methods:

\noindent(1) \textit{Traditional Methods}  encompass recurrent, convolutional, graph, and Transformer-based models. \textbf{GRU4Rec} {(ICLR'16)}~\cite{GRU4Rec}, \textbf{Caser} {(WSDM'18)}~\cite{Caser}, \textbf{HGN} {(KDD'19)}~\cite{HGN}, and \textbf{NextItNet} {(WSDM'19)}~\cite{NextItNet} capture sequential dependencies using GRUs, CNNs, hierarchical gating, and dilated convolutions, respectively. \textbf{LightGCN} {(SIGIR'20)}~\cite{LightGCN} simplifies graph convolution to model high-order connectivity. \textbf{SASRec} {(ICDM'18)}~\cite{SASRec} and \textbf{BERT4Rec} {(CIKM'19)}~\cite{BERT4Rec} employ uni- and bi-directional self-attention mechanisms to learn discriminative user representations.

\noindent(2) \textit{Generative Methods} formulate recommendation as a sequence generation task. \textbf{TIGER} {(NeurIPS'23)}~\cite{TIGER} utilizes RQ-VAE to discretize items into hierarchical SIDs for autoregressive prediction. \textbf{LETTER} {(CIKM'24)}~\cite{LETTER} extends TIGER by integrating collaborative and diversity regularization into RQ-VAE to improve the quality of learned SIDs. \textbf{EAGER} {(KDD'24)}~\cite{EAGER} employs a dual-stream architecture to synergize behavioral and semantic tokens via confidence-based ranking. \textbf{ActionPiece} {(ICML'25)}~\cite{ActionPiece} performs context-aware tokenization, fusing user actions with item content to model fine-grained interest evolution.

\noindent\textbf{Evaluation Metrics.}
We adopt the standard leave-one-out~\cite{elisseeff2003leave} evaluation protocol and rank the ground-truth target item against the whole item set.
% For each user's interaction history, the last item is held out for testing, the second-to-last item for validation, and the rest of the items for training. 
We report performance using two top-K ranking metrics: Recall@K and Normalized Discounted Cumulative Gain (NDCG)@K, with K taking values from \{10, 20\}.

\noindent\textbf{Implementation Details.} 
Traditional baselines utilize RecBole~\cite{zhao2021recbole}. Generative methods share a 4-layer Transformer Encoder-Decoder ($d_{model}\!=\!128$, FFN dim $d_{\mathrm{ff}}\!=\!1024$, 6 heads). ActionPiece requires a larger backbone on CDs ($d_{model}\!=\!256$, $d_{\mathrm{ff}}\!=\!2048$) following its original setup, resulting in $\sim$23.4M parameters compared to PRISM's $\sim$5.5M. PRISM employs 3-layer SIDs (codebook size 256, $d_{cb}\!=\!32$) and MoE ($d_{moe}\!=\!256$, 3 experts, top-2 routing). Training uses Adam (lr=$5 \!\times\! 10^{-4}$, batch 128, 300 epochs) on NVIDIA RTX 4090 GPUs. Hyperparameters are set to $\beta\!=\!0.25$~\cite{TIGER, LETTER}, $\lambda_1\!=\!0.8$, $\lambda_2\!=\!0.2$, and $\gamma\!=\!5 \!\times\! 10^{-4}$. To ensure robustness without extensive tuning, module-specific parameters act as structural constraints: gating margin $\delta\!=\!0.1$ prevents variance collapse, and HSA temperature $\tau_{hsa}\!=\!0.15$ regulates attention sharpness. For ATS, we adopt a fixed range $[\tau_{\min}, \tau_{\max}]\!=\![0.5, 1.0]$ with sensitivity $\alpha\!=\!0.5$, where $N_{ref}\!\approx\!\sqrt{|\mathcal{I}|}/2$ is designed to dynamically adapt to the dataset size.

% \vspace{-10pt}
\subsection{Overall Performance Comparison}\label{sec:main_results}
Table~\ref{tab:main_results} summarizes the overall performance.
We observe that generative models generally outperform traditional ones, confirming the effectiveness of the generative paradigm in capturing complex sequential dependencies. 
Across all datasets and metrics, PRISM achieves the best results, attaining optimal performance on 15 out of 16 cases, % with an average improvement of 11.72\%, 
followed by ActionPiece and TIGER. 
In particular, on the sparser and larger CDs dataset, PRISM improves Recall@10 by 33.9\% over TIGER. 
Moreover, while ActionPiece is competitive on smaller datasets, it requires a $4\times$ larger backbone to attain competitive results on the CDs. Even against such a significant parameter advantage, PRISM still surpasses ActionPiece using a smaller model parameter scale. This indicates that simple model parameter scaling cannot substitute for effective semantic alignment. In addition, hybrid quantization baselines, such as LETTER and EAGER, remain constrained by the inherent information loss introduced during the quantization process. In contrast, PRISM overcomes this through its integrated semantic recommender, which adaptively compensates for quantization loss during inference. 
These results collectively demonstrate the effectiveness of PRISM, even in high-sparsity scenarios where other methods underperform.

\subsection{Robustness to Data Sparsity}
\label{sec:longtail}

\begin{figure}[t]
    \centering
    % 调整图片上方的间距
    % \vspace{-2mm}
    \includegraphics[width=\linewidth]{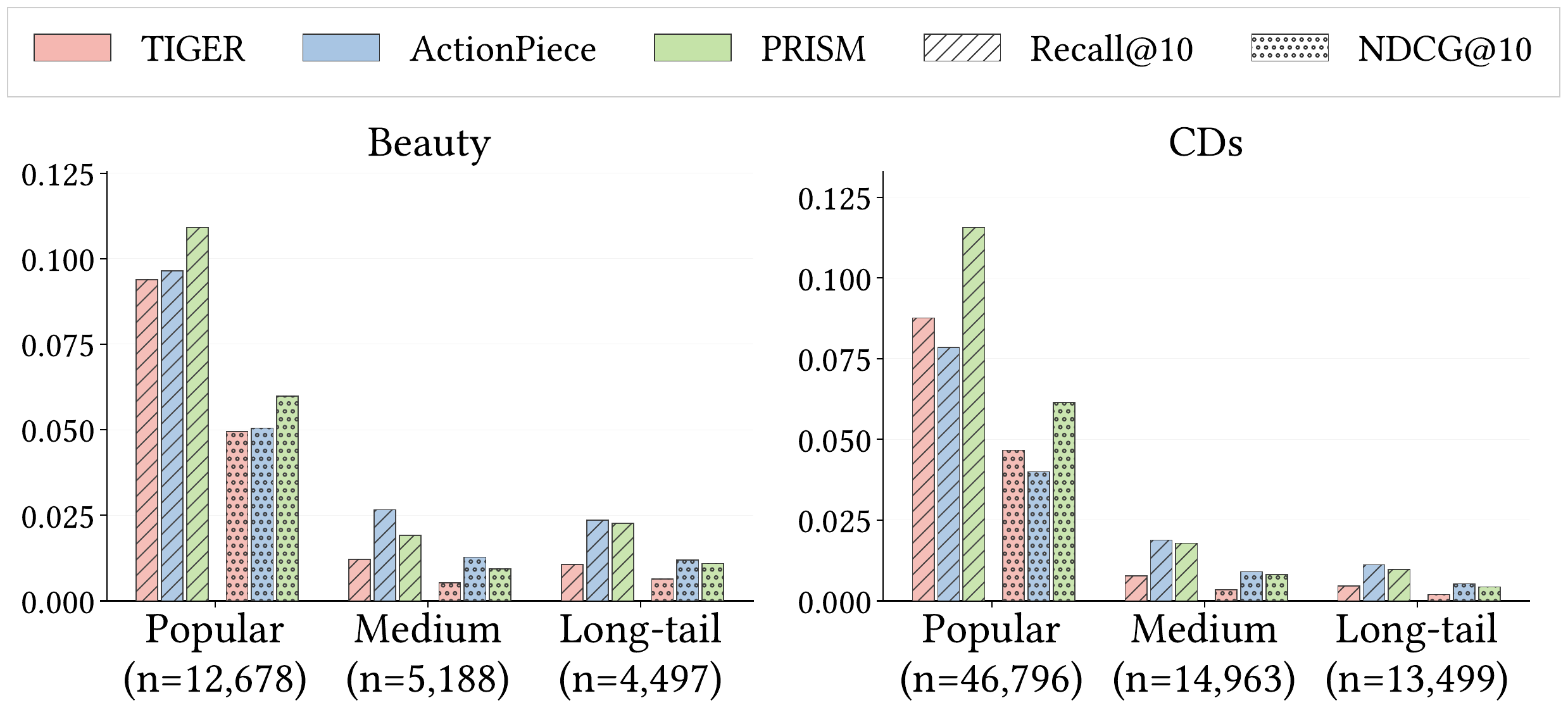}     
    % 调整 Caption 间距
    \vspace{-18pt}
    \Description{A grouped bar chart comparing the performance of TIGER, ActionPiece, and PRISM models across three item popularity groups: Popular, Medium, and Long-tail. The chart displays results for two datasets: Beauty (left) and CDs (right). The Y-axis represents Recall@10 and NDCG@10 scores. The bars show that PRISM consistently outperforms the baselines, with the most significant performance gap observed in the Medium and Long-tail categories on the CDs dataset, indicating superior robustness to data sparsity.}
    \caption{Performance comparison across item popularity groups, where `n' denotes the number of test interactions.}
    \label{fig:longtail}
    % 调整与正文的间距
    \vspace{-12pt}
\end{figure}

To evaluate the robustness of PRISM to data sparsity, we take the Beauty and CDs datasets as examples, and divide target items into Popular, Medium, and Long-tail groups based on their interaction frequency. Note that each group has equal items but highly imbalanced interactions.
As shown in Figure~\ref{fig:longtail}, the baseline TIGER exhibits a sharp performance decline on long-tail items. This limitation stems from severe codebook collapse within TIGER, where the quantization process fails to assign distinct identifiers to low-frequency items, causing their unique semantic patterns to be overshadowed by popular codes. 
In contrast, both PRISM and ActionPiece achieve substantial improvements in these sparse regions. For instance, on the CDs dataset, their performance on both medium-frequency and long-tail items is more than doubled compared to TIGER. These results confirm that constructing distinguishable SIDs and employing context-aware modeling effectively compensates for insufficient behavioral data, enabling the models to learn high-quality embeddings even for low-frequency items.

Comparatively, ActionPiece exhibits a slight advantage on long-tail items. This gain primarily derives from its dynamic tokenization mechanism, which allows features from popular codes to be shared with long-tail codes, and from its use of a backbone that is $4\times$ larger, thereby strengthening its capacity to memorize long-tail information. 
However, this strategy comes at the cost of degraded performance on popular items, with a clear drop observed in the popular group of CDs. PRISM achieves a superior Pareto balance on this issue, delivering improvements on long-tail items comparable to ActionPiece while comprehensively outperforming it on popular items. 
Consequently, PRISM attains the best overall performance on both datasets, indicating that it can effectively cope with data sparsity without compromising recommendation quality for the majority of users through its robust integrated semantic modeling.

\subsection{Ablation Study}\label{sec:ablation}

\begin{table}[t]
\centering
% \small
\vspace{-10pt}
\caption{Analysis of SIDs quality on Beauty.}
\label{tab:sid_quality}
\vspace{-7pt}
\setlength{\tabcolsep}{5pt}
\renewcommand{\arraystretch}{0.9}
\begin{tabular}{l ccc}
\toprule
\multirow{2}{*}{\textbf{Method}} & \multicolumn{2}{c}{\textbf{Collision Rate} ($\downarrow$)} & \multirow{2}{*}{\textbf{Perplexity} ($\uparrow$)} \\
\cmidrule(lr){2-3}
 & Layer 2 & Final & \\
\midrule
% --- Baselines ---
TIGER & 95.73\% & 31.57\% & 84.2 \\
EAGER & - & \textbf{0.00\%} & - \\
ActionPiece & 56.33\% & 16.20\% & 231.5 \\
LETTER & 17.59\% & 0.42\% & 194.1 \\
\midrule
% --- Ablations & PRISM ---
PRISM (w/o HSA) & 28.67\% & 4.77\% & 210.3 \\
PRISM (w/o ACD) & 19.85\% & 2.78\% & 241.8 \\
PRISM (w/o DHR) & 18.22\% & 2.21\% & 245.2 \\
\rowcolor{gray!15} \textbf{PRISM (Full)} & \textbf{17.57\%} & 1.79\% & \textbf{248.5} \\
\bottomrule
\end{tabular}
% }
\vspace{-10pt}
\end{table}

% \begin{table}[t]
% \centering
% % \footnotesize 
% \vspace{-10pt}
% \caption{Analysis of SIDs quality on Beauty.}
% \label{tab:sid_quality}
% \vspace{-7pt}
% \setlength{\aboverulesep}{0pt}
% \setlength{\belowrulesep}{0pt}
% \setlength{\tabcolsep}{4pt} % 稍微收紧列宽
% \renewcommand{\arraystretch}{0.9} 
% \begin{tabular}{l ccc}
% \toprule
% \multirow{2}{*}{\textbf{Method}} & \multicolumn{2}{c}{\textbf{Collision Rate} ($\downarrow$)} & \multirow{2}{*}{\textbf{Perplexity} ($\uparrow$)} \\
% \cmidrule(lr){2-3}
%  & Layer 2 & Final & \\
% \midrule
% % --- Baselines ---
% TIGER & 95.73\% & 31.57\% & 84.2 \\
% EAGER & - & \textbf{0.00\%} & - \\
% ActionPiece & 56.33\% & 16.20\% & 231.5 \\
% LETTER & 17.59\% & 0.42\% & 194.1 \\
% \midrule
% % --- Ablations & PRISM ---
% PRISM (w/o HSA) & 28.67\% & 4.77\% & 210.3 \\
% PRISM (w/o ACD) & 19.85\% & 2.78\% & 241.8 \\
% PRISM (w/o DHR) & 18.22\% & 2.21\% & 245.2 \\
% \rowcolor{gray!15} \textbf{PRISM (Full)} & \textbf{17.57\%} & 1.79\% & \textbf{248.5} \\
% \bottomrule
% \end{tabular}
% \vspace{-10pt}
% \end{table}

\begin{table}[t]
\centering
\caption{Ablation study on Beauty.}
% \small
\label{tab:ablation_rec}
\vspace{-7pt}
\renewcommand{\arraystretch}{1}
\begin{tabular}{cl cccc}
\toprule
\multicolumn{2}{l}{Variant} & R@10 & R@20 & N@10 & N@20 \\
\midrule
% Full Model (跨两列)
\multicolumn{2}{l}{\cellcolor{gray!10}\textbf{PRISM (Full)}} & 
\cellcolor{gray!10}\textbf{0.0713} & 
\cellcolor{gray!10}\textbf{0.1030} & 
\cellcolor{gray!10}\textbf{0.0387} & 
\cellcolor{gray!10}\textbf{0.0467} \\
\midrule
% Stage 1
\multirow{3}{*}{\shortstack[l]{{Tokenization} }} 
 & w/o ACD & 0.0691 & 0.1029 & 0.0375 & 0.0461 \\
 & w/o HSA & 0.0652 & 0.1001 & 0.0341 & 0.0430 \\
 & w/o DHR & 0.0688 & 0.1015 & 0.0368 & 0.0451 \\
\cmidrule{1-6} 
% Stage 2
\multirow{3}{*}{\shortstack[l]{{Generation} }} 
 & w/o DSI & 0.0675 & 0.0968 & 0.0366 & 0.0439 \\
 & w/o SSA & 0.0695 & 0.1021 & 0.0379 & 0.0460 \\
 & w/o ATS & 0.0682 & 0.1007 & 0.0363 & 0.0445 \\
\bottomrule
\end{tabular}
\vspace{-12pt}
\end{table}

% \begin{table}[t]
% \centering
% \caption{Ablation study on Beauty.}
% \vspace{-7pt}
% \label{tab:ablation_rec}
% % \small 
% \setlength{\tabcolsep}{4pt}
% \renewcommand{\arraystretch}{0.95} % 压缩行高
% \setlength{\aboverulesep}{0pt} % 消除横线周围间距
% \setlength{\belowrulesep}{0pt}
% \begin{tabular}{ll cccc}
% \toprule
% \multicolumn{2}{l}{\textbf{Variant}} & R@10 & R@20 & N@10 & N@20 \\
% \midrule
% \rowcolor{gray!15} 
% \multicolumn{2}{l}{\textbf{PRISM (Full)}} & \textbf{0.0713} & \textbf{0.1030} & \textbf{0.0387} & \textbf{0.0467} \\
% \midrule
% \multirow{3}{*}{\textbf{Tokenization}} 
%  & w/o ACD & 0.0691 & 0.1029 & 0.0375 & 0.0461 \\
%  & w/o HSA & 0.0652 & 0.1001 & 0.0341 & 0.0430 \\
%  & w/o DHR & 0.0688 & 0.1015 & 0.0368 & 0.0451 \\
% \midrule 
% \multirow{3}{*}{\textbf{Generation}} 
%  & w/o DSI & 0.0675 & 0.0968 & 0.0366 & 0.0439 \\
%  & w/o SSA & 0.0695 & 0.1021 & 0.0379 & 0.0460 \\
%  & w/o ATS & 0.0682 & 0.1007 & 0.0363 & 0.0445 \\
% \bottomrule
% \end{tabular}
% \vspace{-10pt}
% \end{table}

\subsubsection{Quality Analysis of SIDs}

To assess the quality of the learned SIDs, we evaluate the metrics listed in Table~\ref{tab:sid_quality}. 
The collision rate (CR) at intermediate layers prior to deduplication reflects coarse-grained semantic similarity, while the collision rate in the final layer indicates the model's ultimate discriminative capability. Codebook perplexity (PPL) quantifies the uniformity of token distribution, where a value closer to codebook size of 256 indicates more balanced utilization of the latent space.

We observe that, except for PRISM, other generative methods struggle to balance collision rate and codebook utilization. 
TIGER suffers from severe codebook collapse, with a PPL of only 84.2 and a CR as high as 31.57\%. 
Although ActionPiece, which uses Optimized Product Quantization~\cite{OPQ}, achieves a relatively high PPL of 231.5, its lack of structural constraints still results in a CR as high as 16.20\%. 
EAGER achieves zero CR through hard K-means clustering, but it fails to effectively fuse heterogeneous modalities. 
LETTER significantly reduces CR, but severely underutilizes the capacity of the codebook, with a PPL of only 194.1, which is clearly suboptimal.

In contrast, PRISM achieves the best balance between these two aspects, reaching a near-optimal PPL of 248.5 while keeping CR down to just 1.79\%. This indicates that PRISM can fully and uniformly activate the codebook space, and the ablation results further validate our design. 
Removing HSA causes PPL to drop to 210.3 and the CR to rise to 4.77\%. More critically, the CR at the second layer soars to 28.67\%, demonstrating that hierarchical anchors are essential for regularizing the tree structure and preventing collapse in intermediate layers. Furthermore, removing ACD or DHR likewise degrades performance on both metrics, confirming their roles in refining quantization boundaries and preserving feature topology.

\subsubsection{Impact of Modules on Recommendation}
We conduct an additional ablation study to evaluate the impact of each key module in PRISM, as reported in Table~\ref{tab:ablation_rec}. 
In the semantic tokenization stage, the HSA module is critical for structural stability. Removing it causes Recall@10 to drop sharply to 0.0652, as the codebook latent space loses its hierarchical structure without semantic anchors. DHR module is equally vital for ranking accuracy, as removing DHR substantially drops NDCG@10 to 0.0368. This confirms that explicitly preserving collaborative signals during quantization is essential for learning clean preferences. In addition, removing ACD also degrades performance, indicating that filtering out interaction noise prevents PRISM from fitting spurious correlations. 

In generative recommendation, removing DSI results in a low Recall@10 of 0.0675, which demonstrates that incorporating continuous features can effectively compensate for the information loss of discrete SIDs. 
The performance drop without ATS supports our hypothesis that 
varying tree densities require dynamic uncertainty calibration. 
Finally, the SSA module aligns the generation process with the true hierarchical structure, effectively mitigating semantic drift during inference and ensuring consistent performance gains.

\begin{figure}[t]
    \centering
    % 调整图片上方的间距
    \vspace{-8pt}
    \includegraphics[width=\linewidth]{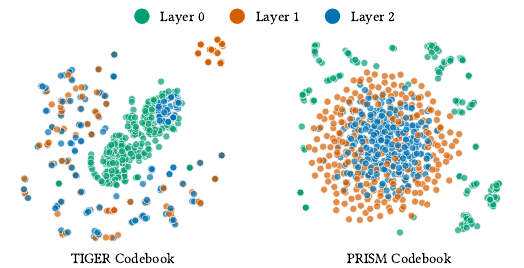}     
    \vspace{-18pt}
    \Description{A composite figure displaying t-SNE visualizations of latent structures. The top row compares codebook embeddings, where TIGER shows a collapsed, unstructured cluster, whereas PRISM exhibits a distinct, concentric ring structure separating different hierarchical layers.}
    \caption{t-SNE visualization of codebook embeddings.}
    \label{fig:tsne_codebooks}    
    \vspace{-13pt}
\end{figure}

\begin{figure}[t]
    \centering
    % 调整图片上方的间距
    % \vspace{-8pt}
    \includegraphics[width=\linewidth]{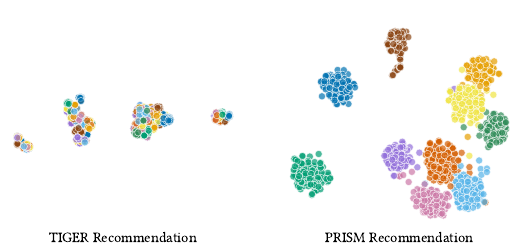}    
    \vspace{-18pt}
    \Description{A composite figure displaying t-SNE visualizations of latent structures. The bottom row compares item embeddings colored by category. TIGER's embeddings show entangled clusters with fuzzy boundaries, while PRISM displays well-separated, compact clusters that align clearly with semantic categories.}
    \caption{t-SNE visualization of item embeddings, where different colors indicate different categories.}
    \label{fig:tsne_recommendations}    
    \vspace{-10pt}
\end{figure}

\begin{figure*}[htbp]
    \centering
    % 设置极小的列间距，让图片紧密排列
    \setlength{\tabcolsep}{0.5pt} 
    
    % --- 1. Layer ---
    \begin{subfigure}[b]{0.163\textwidth}
        \centering
        \includegraphics[width=\linewidth]{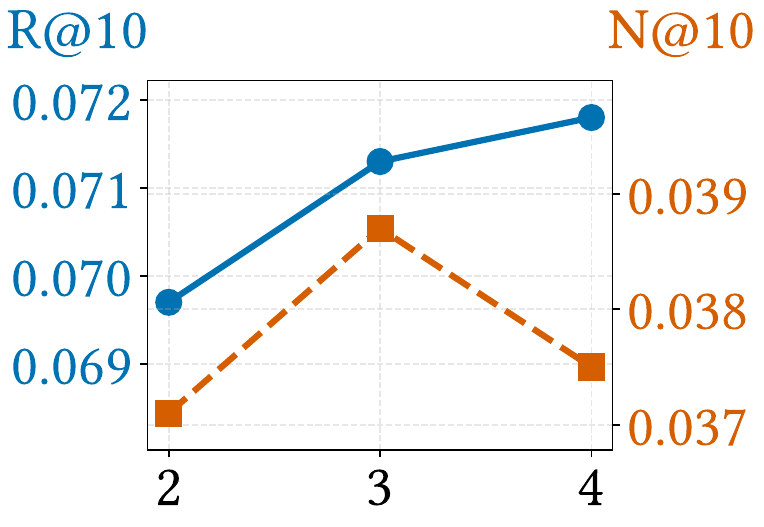}
        \vspace{-5mm} % 调整图片与子标题的间距，可根据需要修改
        \caption{$L$}
        \label{fig:sens_layer}
    \end{subfigure}
    \hfill
    % --- 2. Dim ---
    \begin{subfigure}[b]{0.163\textwidth}
        \centering
        \includegraphics[width=\linewidth]{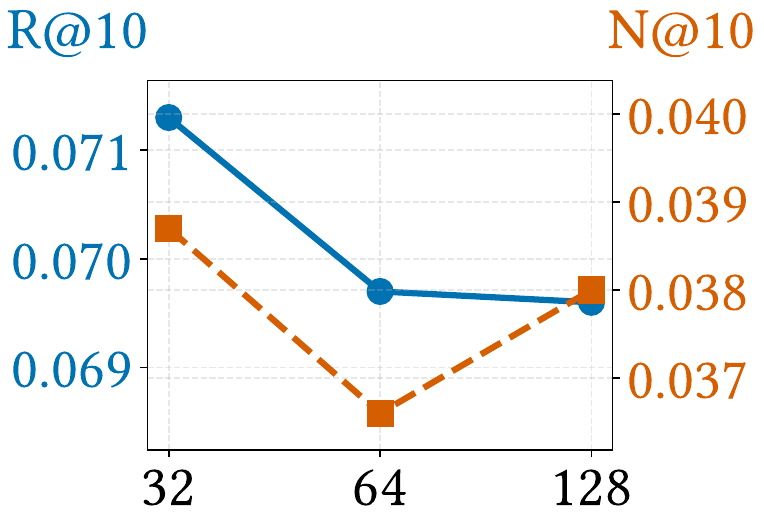}
        \vspace{-5mm}
        \caption{$d_{cb}$}
        \label{fig:sens_dim}
    \end{subfigure}
    \hfill
    % --- 3. ACD ---
    \begin{subfigure}[b]{0.163\textwidth}
        \centering
        \includegraphics[width=\linewidth]{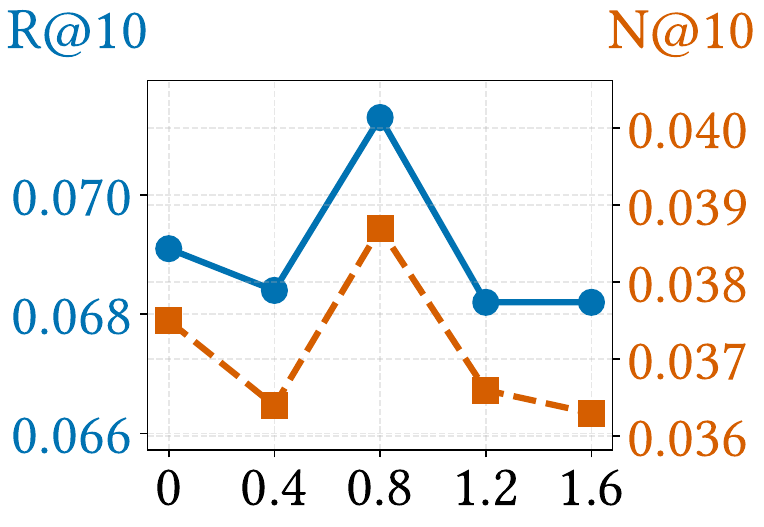}
        \vspace{-5mm}
        \caption{$\lambda_1$}
        \label{fig:sens_acd}
    \end{subfigure}
    \hfill
    % --- 4. HSA ---
    \begin{subfigure}[b]{0.163\textwidth}
        \centering
        \includegraphics[width=\linewidth]{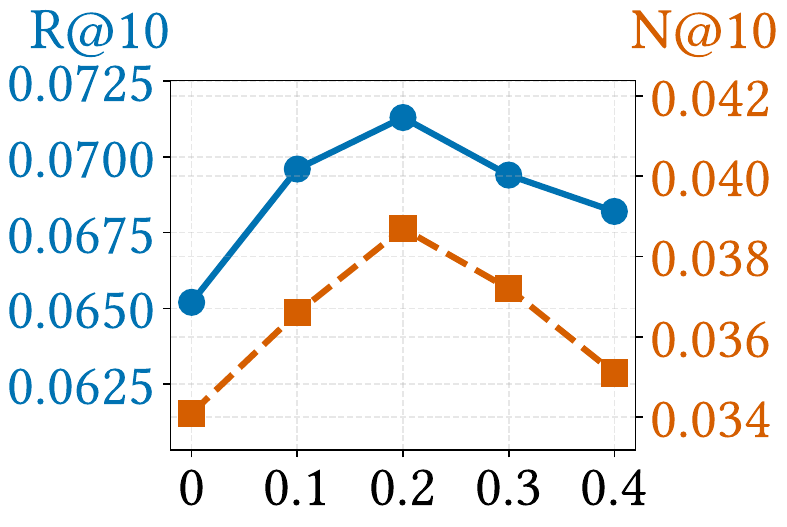}
        \vspace{-5mm}
        \caption{$\lambda_2$}
        \label{fig:sens_hsa}
    \end{subfigure}
    \hfill
    % --- 5. SSA ---
    \begin{subfigure}[b]{0.163\textwidth}
        \centering
        \includegraphics[width=\linewidth]{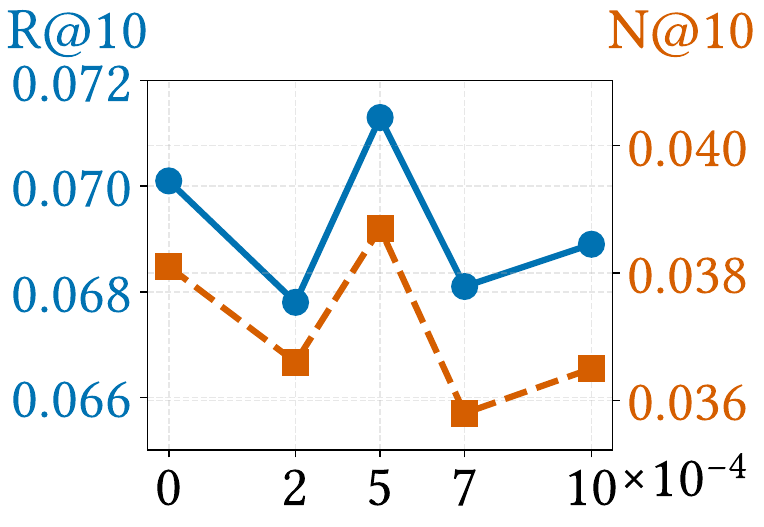}
        \vspace{-5mm}
        \caption{$\gamma$}
        \label{fig:sens_ssa}
    \end{subfigure}
    \hfill
    % --- 6. MoE ---
    \begin{subfigure}[b]{0.163\textwidth}
        \centering
        \includegraphics[width=\linewidth]{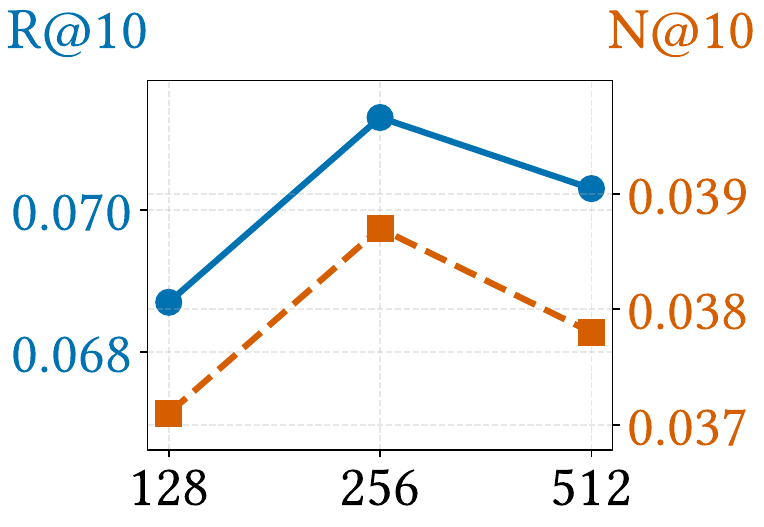}
        \vspace{-5mm}
        \caption{$d_{moe}$}
        \label{fig:sens_moe}
    \end{subfigure}

    \vspace{-3mm} % 压缩子标题与主Caption的距离
    \Description{A set of six line charts labeled (a) through (f) analyzing hyperparameter sensitivity on the Beauty dataset. The charts plot Recall@10 (blue solid line, left axis) and NDCG@10 (orange dashed line, right axis) against varying values of six parameters: codebook depth L, embedding dimension d_cb, regularization weights lambda_1, lambda_2, and gamma, and MoE dimension d_moe. The trends generally show a bell-shaped curve, indicating optimal performance at intermediate values for most parameters.}
    \caption{Hyper-parameter sensitivity analysis on the Beauty dataset. We visualize the trade-off between Recall@10 (blue circles, left axis) and NDCG@10 (orange squares, right axis) by varying one parameter while fixing others.}
    \label{fig:hyper_sensitivity}
    \vspace{-5pt}
\end{figure*}

\subsection{Qualitative Visualization of Latent Structure}

To visualize PRISM's structural superiority, we use t-SNE to visualize the embeddings from both tokenization and recommendation stages. 
Each item embedding is derived via average pooling of the embeddings of corresponding SIDs. We primarily compare PRISM against TIGER, as both employ the residual quantization paradigm, which allows a direct comparison of their manifold structures.

The codebook visualizations in Figure~\ref{fig:tsne_codebooks} reveal codebook collapse in TIGER, which exhibits a degenerate structure, where the majority of embeddings clump into a dense mass. This is consistent with the high collision rate of 31.57\% reported in Table~\ref{tab:sid_quality}, further confirming that the standard reconstruction objective fails to prevent codebook collapse in RQ-VAE. In sharp contrast, PRISM displays a distinct concentric distribution, with different codebook layers forming well-separated ring-like structures. The outermost ring shows a pronounced clustering pattern, indicating that the first layer effectively captures coarse-grained global categories. Moving inward, the distribution becomes more uniform and no longer exhibits strong clusters, suggesting that deeper layers focus on modeling fine-grained residuals to refine item representations. This ordered structural organization indicates that PRISM successfully regularizes the latent space, forcing the model to distinguish semantic hierarchies and maximize codebook utilization.

We further examine how the synergy of structured quantization and dynamic semantic integration translates to item discriminability, as shown in Figure~\ref{fig:tsne_recommendations}, where item embeddings from recommendation backbones are colored by different categories. 
TIGER presents clusters with blurred boundaries and entanglement, indicating failure to capture clear semantic boundaries. 
Conversely, PRISM forms compact, well-separated clusters that are aligned with semantic categories. 
This provides additional evidence that  PRISM not only maximizes codebook utilization but also successfully injects high-level category logic into discrete identifiers, driving superior recommendation performance.

\vspace{-5pt}
\subsection{Hyperparameter Sensitivity}\label{sec:hyperparams}

We investigate hyperparameter sensitivity on the Beauty dataset. Figure~\ref{fig:hyper_sensitivity} visualizes the trade-off between Recall@10 and NDCG@10 by varying one parameter with others fixed.

\textbf{Codebook Structure ($L, d_{cb}$).}
We first analyze the semantic indexing structure by varying the codebook depth $L \in \{2, 3, 4\}$ and embedding dimension $d_{cb} \in \{32, 64, 128\}$. Figure~\ref{fig:sens_layer} shows that increasing the depth from 2 to 3 yields significant gains, suggesting that a 3-layer hierarchy is necessary to capture fine-grained semantic distinctions. However, further increasing $L$ to 4 leads to diminishing returns. This is likely because deeper hierarchies generate longer ID sequences, which accumulates errors during generation. 
For the codebook dimension, Figure~\ref{fig:sens_dim} shows that a compact dimension of $d_{cb}=32$ achieves the best performance. Unlike continuous embeddings, which typically benefit from higher dimensions, discrete codes gain from low-dimensional, compact embeddings that enforce information compression.

\textbf{Regularization Strengths ($\lambda_1, \lambda_2, \gamma$).}
We examine the contribution of auxiliary objectives by sweeping weights $\lambda_1$ (ACD), $\lambda_2$ (HSA), and $\gamma$ (SSA) from 0. Figures~\ref{fig:sens_acd}-\ref{fig:sens_ssa} show bell-shaped trends. 
Crucially, the performance at a weight of 0 is consistently lower than the peak performance, which is achieved at moderate weight values ($\lambda_1=0.8, \lambda_2=0.2, \gamma=5 \!\times\! 10^{-4}$).
Once the weights exceed these thresholds, we observe a decline across the metrics, indicating that overly large regularization weights can overshadow the primary reconstruction objective, leading to suboptimal representation learning of the intrinsic item semantics.

\textbf{MoE Capacity ($d_{moe}$).}
Finally, we evaluate the capacity of the MoE fusion module by varying the expert hidden dimension $d_{moe} \in \{128, 256, 512\}$. As shown in Figure~\ref{fig:sens_moe}, increasing the dimension from 128 to 256 improves MoE capacity. However, further expanding $d_{moe}$ to 512 causes slight degradation. Given sparse interactions in the dataset, an over-parameterized fusion layer tends to overfit the training noise. Therefore, we adopt $d_{moe}=256$ to balance expressiveness and generalization.

\subsection{Efficiency Analysis}\label{sec:efficiency}

\begin{figure}[!t]
    \centering
    % 调整图片上方的间距
    % \vspace{-2mm}
    \includegraphics[width=\linewidth]{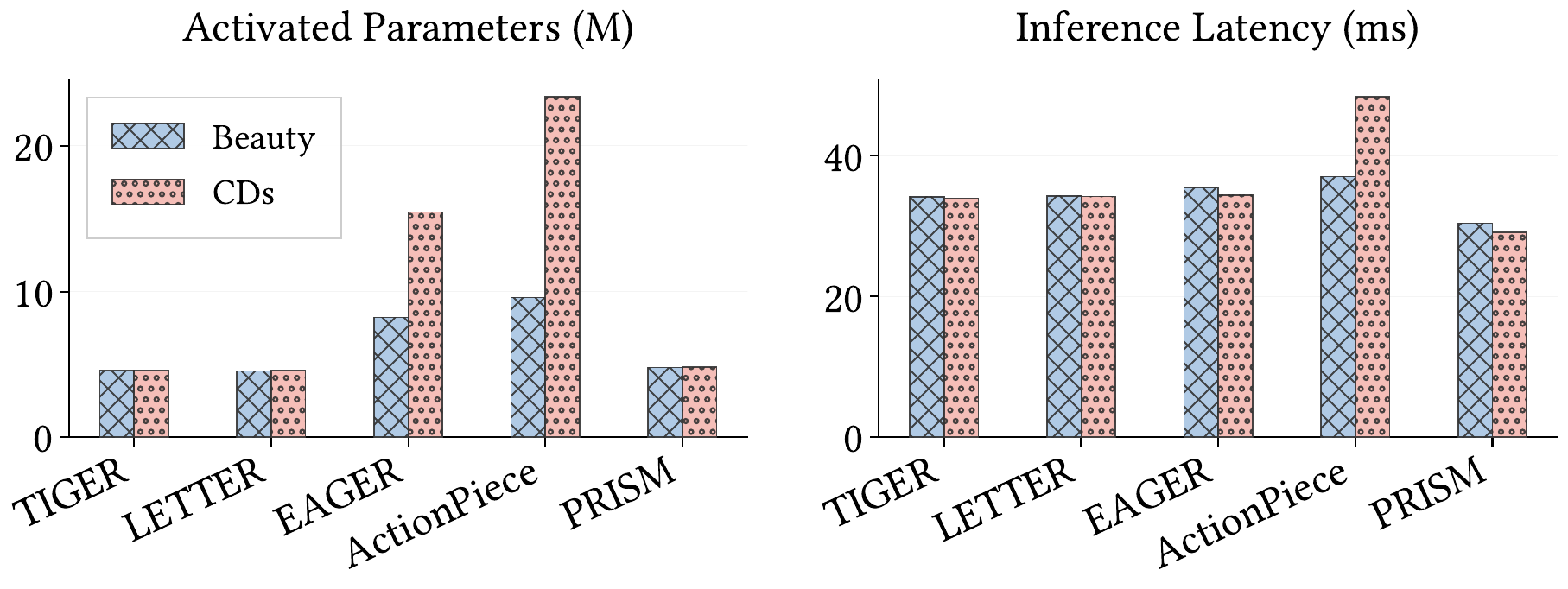}     
    % 调整 Caption 间距
    \vspace{-8mm}
    \Description{Two bar charts comparing the efficiency of five generative models: TIGER, LETTER, EAGER, ActionPiece, and PRISM. The left chart shows Activated Parameters in millions, and the right chart shows Inference Latency in milliseconds. Data is provided for Beauty and CDs datasets. PRISM demonstrates a highly efficient profile with low parameter count (around 5.5M) and low latency (under 30ms) on both datasets, whereas ActionPiece shows a drastic increase in both parameters and latency on the larger CDs dataset.}
    \caption{Comparison of activated parameters and inference latency across generative methods on Beauty and CDs.}
    \label{fig:efficiency}    
    % 调整与正文的间距
    \vspace{-10pt}
\end{figure}

We evaluate efficiency on Beauty and CDs to assess the trade-off between capacity and speed. Given the superior performance of generative methods in Table~\ref{tab:main_results} and their inherent scalability, we focus exclusively on this paradigm. 
As shown in Figure~\ref{fig:efficiency}, despite CDs being $5\times$ larger than Beauty (64K vs. 12K), generative models maintain stable latency and parameter usage. This stability makes the framework suitable for industrial applications with rapidly scaling catalogs. 
Among them, PRISM achieves the best balance between efficiency and performance. ActionPiece incurs the highest overhead on the large-scale CDs, with 23.4M activated parameters and an inference latency of 48.4ms. This is due to its need to employ a backbone network four times larger to maintain sufficient capacity for sparse data. 
For EAGER, although its parameter count on CDs nearly doubles to 15.5M due to the expansion of the embedding table, its inference latency remains stable at around 34ms, since its computational cost is mainly determined by the fixed length of the code sequence rather than the size of the item pool. 
In contrast, PRISM exhibits stronger adaptability, maintaining a compact parameter size of only 5.5M on CDs and achieving the lowest latency of 29.1ms. This efficiency stems from the sparse MoE mechanism, which scales up model capacity without proportionally increasing inference latency, and high-quality SIDs that enable a lightweight backbone without sacrificing performance.

\section{Conclusion}
\label{sec:conclusions}
In this paper, we identify and address two critical limitations hindering lightweight generative sequential recommendation: impure and unstable semantic tokenization, and lossy and weakly structured generation. To address these limitations, we propose PRISM, a unified framework that synergizes purified structural tokenization with dynamic semantic integration. Specifically, we introduce the Purified Semantic Quantizer, which constructs a robust codebook by filtering interaction noise via adaptive collaborative denoising and enforcing structural stability through hierarchical semantic anchoring. Building upon these purified representations, we design the Integrated Semantic Recommender that effectively mitigates quantization loss by employing dynamic semantic integration to fuse fine-grained continuous features, while ensuring logical correctness via semantic structure alignment. Extensive experiments on four real-world datasets demonstrate that PRISM significantly outperforms state-of-the-art baselines, exhibiting exceptional robustness, particularly in highly challenging data-sparse scenarios.
% For future work, we plan to explore the integration of richer multi-modal content, such as visual features, to further enhance the semantic granularity of the quantizer and recommender. Additionally, we plan to explore efficiency-aware mechanisms to leverage the vast semantic capabilities of LLMs, ensuring superior reasoning depth while retaining the low latency of our lightweight architecture.

% \balance
\bibliographystyle{ACM-Reference-Format}
\bibliography{9bibfile}

% \newpage
% \appendix
% \input{6-appendix}

\end{document}